\documentclass{icrc2009}

\usepackage{graphicx}   
\usepackage[caption=false]{caption}    
\usepackage[font=footnotesize]{subfig} 
\usepackage{fixltx2e}
\usepackage{url}

\newcommand{\shorttitle}[1]%
{\markboth{Proceedings of the 31\MakeLowercase{$^{st}$} ICRC, {\L}\'{o}d\'{z} 2009}{#1} }
\newcommand{\etal}{\MakeLowercase{\textit{et al. }}} 

\begin{document}
\title{A new strategy to select candidate blazars for VHE observation}

\author{\IEEEauthorblockN{Bagmeet Behera\IEEEauthorrefmark{1}\IEEEauthorrefmark{2} and
			  Stefan J. Wagner\IEEEauthorrefmark{1}}
                            \\
\IEEEauthorblockA{\IEEEauthorrefmark{1}Landessternwarte, Zentrum f\"ur Astronomie der Universit\"at Heidelberg, D-69117, Germany}
\IEEEauthorblockA{\IEEEauthorrefmark{2}Fellow of the International Max Planck Research School for Astronomy and Cosmic Physics\\
at the University of Heidelberg}
}

\shorttitle{Behera \etal Selecting candidate VHE blazars}
\maketitle

\begin{abstract}
A new strategy to select candidate blazars for VHE observations, that gives an inherently complete sample from an all-sky compilation (excluding a narrow region around the galactic plane) of GeV-bright blazars is presented.\\
The strategy is based on selecting candidate VHE blazars from a GeV-detected blazar sample, namely the LAT-bright-AGN-sample (LBAS) of the Fermi Gamma-ray Space Telescope (FGST). The intrinsic spectra in the VHE regime is estimated based on the extrapolation of the GeV spectra measured with FGST. This is corrected for the unavoidable extinction due to the Extragalactic Background Light (EBL) in the UV to mid-IR. The resulting observable-spectra ranked by the integrated flux is used to extract candidates for observations in the VHE regime with Imaging Atmospheric Cherenkov Telescopes (IACTs). The predictions are cross-checked using the sample of known VHE blazars.\\
\end{abstract}

 
\section{Introduction}
The flux level of most blazars pose a challenge to the sensitivity of current IACT experiments. Thus relatively long exposure times are required to detect blazars in the VHE regime. In addition IACTs have a small field of view compared to GeV gamma-ray satellite telescopes. An all sky survey in VHE with a deep exposure is thus not technically feasible using current installations. Therefore it is necessary to adopt some selection procedure to choose candidate VHE blazars to be observed with IACT experiments. A commonly applied strategy to select VHE-observation candidates uses sources selected on the basis of their brightness at lower energies, viz. in the radio and X-ray regime \cite{CostamanteGhisellini2002}. This method relies on the hypothesis that the radio and X-ray frequencies, which sample the synchrotron regime, are a direct proxy to estimate the brightness of the VHE component. This strategy has been quite successful. However, the physical mechanism that connects the separate spectral components is not firmly established. In addition, while applying this strategy, sources were selected from incomplete source-catalogs and were thus an incomplete sample of predicted-VHE-bright blazars.\\

An alternate method proposes to select blazars based on their spectrum in the $100$\,MeV - $100$\,GeV band (from now on, referred to as the GeV-band). This regime is probed using satellite-borne $\gamma$-ray-instruments such as the EGRET instrument which flew on board the CGRO satellite, and the Large Area Telescope (LAT) instrument on the recently launched FGST. For most blazars it is believed that the VHE emission and the GeV emission arise from the same physical mechanism and hence should be intimately related. Thus selecting blazars detected in the GeV-band with certain spectral characteristics, can be a useful tool in selecting candidates for VHE observations. This method was applied to the archival EGRET data from the 3$^{rd}$ EGRET catalog \cite{Hartman1999} in \cite{BeheraWagner2008}. Here we apply this strategy to the recently released GeV-bright AGN sample provided by the FGST team, called the LBAS\cite{Abdo20091}.\\
\begin{figure}[ht]
    \centering
    \includegraphics[width=2.2in, angle=-90]{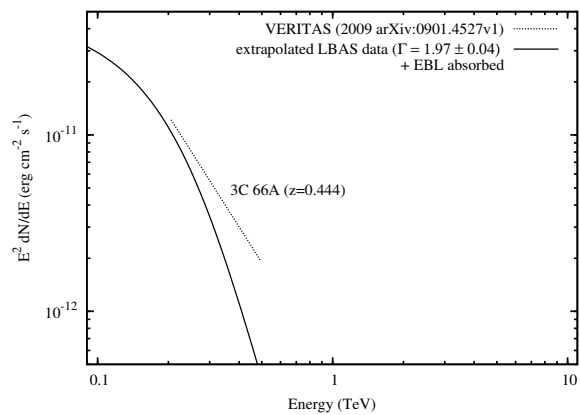}
  \caption{Extrapolated LBAS spectra corrected for EBL attenuation (solid line) is plotted along with a fit to an actual VHE spectral measurement with the VERITAS array.}
    \label{VHEextrapolatiion}
  \end{figure}
\noindent
  \begin{table*}[!th]
  \caption{The list of candidates selected from LBAS.}
  \label{RankedList}
  \centering
  \begin{tabular}{|l|c|r|c|c|c|c|l|}
  \hline
 FGST Name       &  RA    & dec        & z      & F$_{100}$ & $\Gamma_{Fermi}$  & I($>0.2$TeV)   & Other Name \\
		 &        &           &         &           &               &          &      \\
		 &	  &           &         & ($10^{-8}$/cm$^2$/s)&     & ($10^{-12}$/cm$^2$/s) &     \\
   \hline 
0FGL J1104.5+3811  &  166.137 & 38.187 & 0.030 & 15.3 $\pm$  1.1 &  1.77  $\pm$ 0.04 &     349. &    Mrk 421\footnotemark[1]  \\
0FGL J2158.8-3014  &  329.704 & -30.237 & 0.116 & 18.1 $\pm$  1.2 &  1.85 $\pm$  0.04 &  137. &    PKS 2155-304\footnotemark[1]  \\
0FGL J1653.9+3946  &  253.492 & 39.767 &0.033 & 3.1 $\pm$  0.6 &  1.70  $\pm$ 0.09 & 115. &     Mrk 501\footnotemark[1]   \\
0FGL J1053.7+4926  &  163.442 & 49.449 &0.140 & 0.5 $\pm$  0.3 &  1.42 $\pm$  0.20 & 56.6 &  MS1050.7+4946    \\
0FGL J1218.0+3006  &  184.517 &  30.108 &0.130 & 9.7 $\pm$  1.1 &  1.89 $\pm$  0.06 & 51.1 &   B2 1215+30     \\
0FGL J1719.3+1746  &  259.830 & 17.768 &0.137 & 6.9 $\pm$  0.9 &  1.84  $\pm$ 0.07 &  49.5 &  PKS 1717+177    \\
0FGL J1015.2+4927  &  153.809 & 49.463 &0.212 & 4.9 $\pm$  0.7 &  1.73  $\pm$ 0.07 &   47.2 &   1ES 1011+496\footnotemark[1]    \\
0FGL J2000.2+6506  &  300.053 & 65.105 &0.047 & 4.2 $\pm$  1.0 &  1.86 $\pm$  0.11 &    44.6 &   1ES 1959+650\footnotemark[1] \\
0FGL J1221.7+2814  &  185.439  & 28.243 &0.102 & 8.3 $\pm$  1.1 &  1.93 $\pm$  0.07 &  38.6 &    W Comae\footnotemark[1] \\
0FGL J2009.4-4850  &  302.363 & -48.843 &0.071 & 2.9 $\pm$  0.9 &  1.85 $\pm$  0.14 &   28.6	&    PKS 2005-489\footnotemark[1] \\
0FGL J0320.0+4131  &  50.000 & 41.524 &0.018 & 22.1 $\pm$  1.9 &  2.17 $\pm$  0.06 &  27.6	&    0316+413     \\
0FGL J1517.9-2423  &  229.496 & -24.395 &0.048 & 4.1$\pm$   1.2 &  1.94$\pm$   0.14 &  24.1	&   AP Lib     \\
0FGL J0449.7-4348  &  72.435 & -43.815 &0.205 & 12.0 $\pm$  1.3 &  2.01 $\pm$  0.06 &  17.6	&    PKS 0447-439    \\
0FGL J0222.6+4302  &  35.653 & 43.043 &0.444 & 25.9  $\pm$ 1.6 &  1.97 $\pm$  0.04 &  8.79	&   3C 66A\footnotemark[1] \\
0FGL J0722.0+7120  &  110.508 & 71.348 &0.310 & 16.4 $\pm$  1.4 &  2.08  $\pm$ 0.05 &   7.52 &   S5 0716+71 \\ 
0FGL J0507.9+6739  &  76.985 & 67.650 &0.416 & 1.7 $\pm$  0.8 &  1.67 $\pm$  0.18 &   5.36 &  1ES 0502+675    \\
0FGL J2202.4+4217  &  330.622 & 42.299 &0.069 & 8.5 $\pm$  1.8 &  2.24 $\pm$  0.12 &      4.9 & BL Lacertae\footnotemark[1] \\ 
0FGL J0303.7-2410  &  45.940 & -24.176&0.260 & 3.8  $\pm$ 0.9 &  2.01  $\pm$ 0.13 & 3.96  &  PKS 0301-243     \\
0FGL J1751.5+0935  &  267.893 & 9.591 &0.322 & 18.4 $\pm$  2.1 &  2.27  $\pm$ 0.07 &  2.04  &  OT 081 \\ 
0FGL J1512.7-0905  &  228.196 & -9.093 &0.360 & 55.8 $\pm$  3.3 &  2.48 $\pm$  0.05 &  1.05  &  PKS 1510-08 \\ 
\hline
  \end{tabular}
\end{table*}
\section{Predicting the VHE flux}
The LBAS, \cite{Abdo20091} consists of $106$ sources which have high-confidence associations with known AGNs. It contains $104$ blazars; of which $57$ are flat spectrum radio quasars (FSRQs), $42$ are BL Lac objects, and $5$ are blazars with uncertain classification, in addition to the two radio galaxies - Centaurus A and NGC 1275. In the LBAS, the integral flux between $0.1$\,GeV and $100$\,GeV and the corresponding spectral index are given for all sources, derived from a power law fit between $0.2$\,GeV and $100$\,GeV. A search on the redshift of all these sources in the available literature yielded redshift values for 92 objects.\\

A subset of these objects after excluding sources with $z\,<\,1.0$ was used as the source-sample for applying our selection strategy. Sources with uncertain redshifts were excluded. The LBAS spectra of the selected sample was extrapolated to VHE energies to get the estimated intrinsic VHE spectra. In the absence of additional details on individual source spectra (e.g. any spectral-break or turn-off) such an extrapolation is a reasonable first order estimate of the VHE spectra. This estimated intrinsic VHE spectra corrected for the EBL attenuation gives the estimated observed-VHE-spectra for each source. VHE $\gamma$-rays are absorbed in the inter-galactic medium via pair-production mechanism, from photon-photon scattering on the extragalactic photon field ($\gamma_{TeV} + \gamma_{EBL} \rightarrow e^+ + e^-$). The optical depth ($\tau$) due to this mechanism, is both a function of redshift, and the photon energies. For each object the corresponding $\tau(E_{TeV},z)$ is calculated using the EBL model in \cite{Aharonian2006a}. The EBL limits in \cite{Aharonian2006a} are derived from actual VHE observations, making certain assumptions about the intrinsic spectra and are hence a reasonable upper limit to the EBL level. The extrapolated spectra is attenuated by a factor of $e^{-\tau}$ to obtain the predicted observed spectra in VHE (see figure \ref{VHEextrapolatiion}). \footnotetext[1]{Already detected at TeV energies} \\
  \begin{figure*}[!ht]
   \centerline{\subfloat[Non simultaneous measurements]{\includegraphics[width=2.2in, angle=-90]{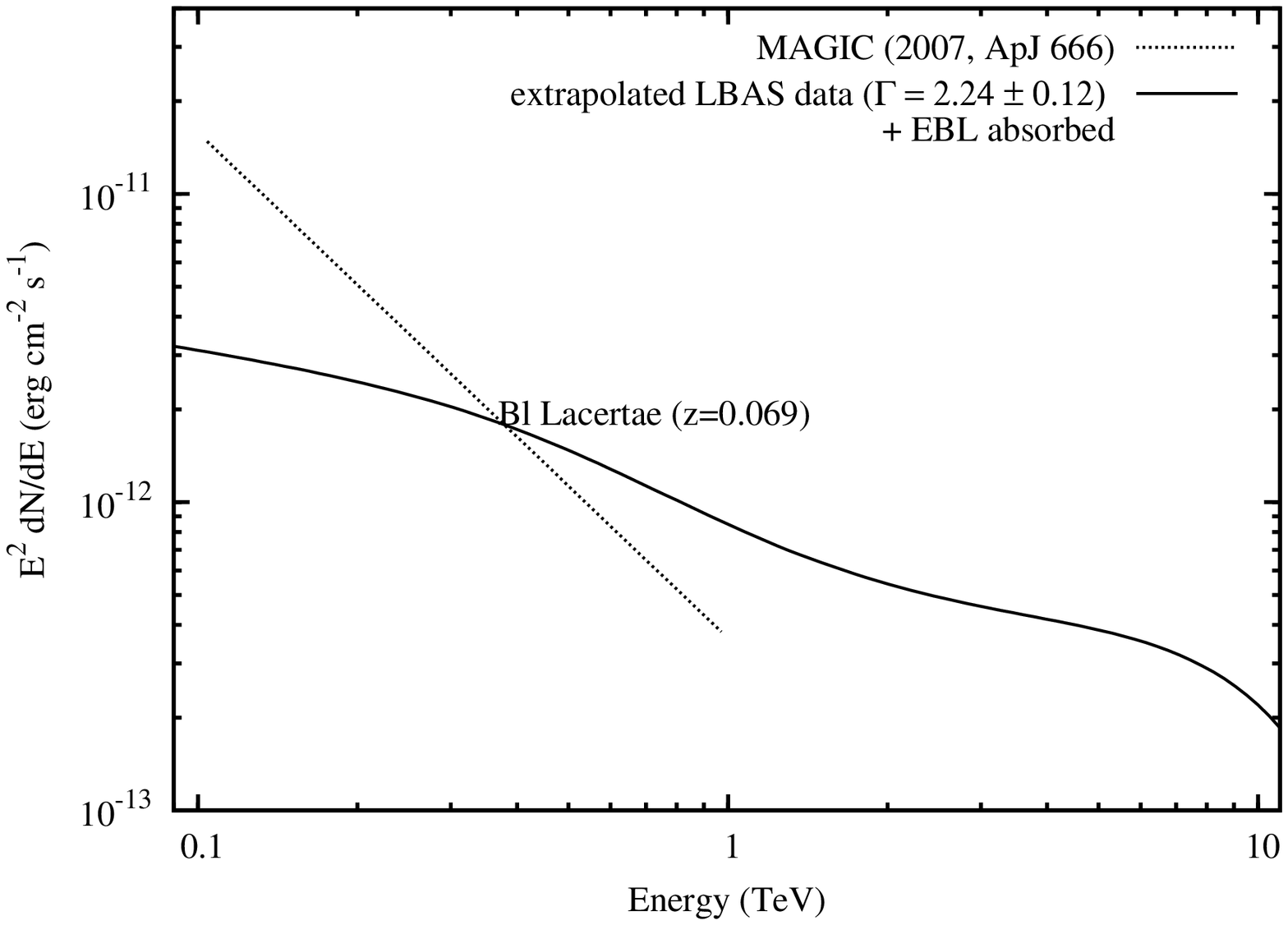}
	  \label{Bl-Lac}}
              \hfil
              \subfloat[strictly simultaneous measurements]{\includegraphics[width=2.2in, angle=-90]{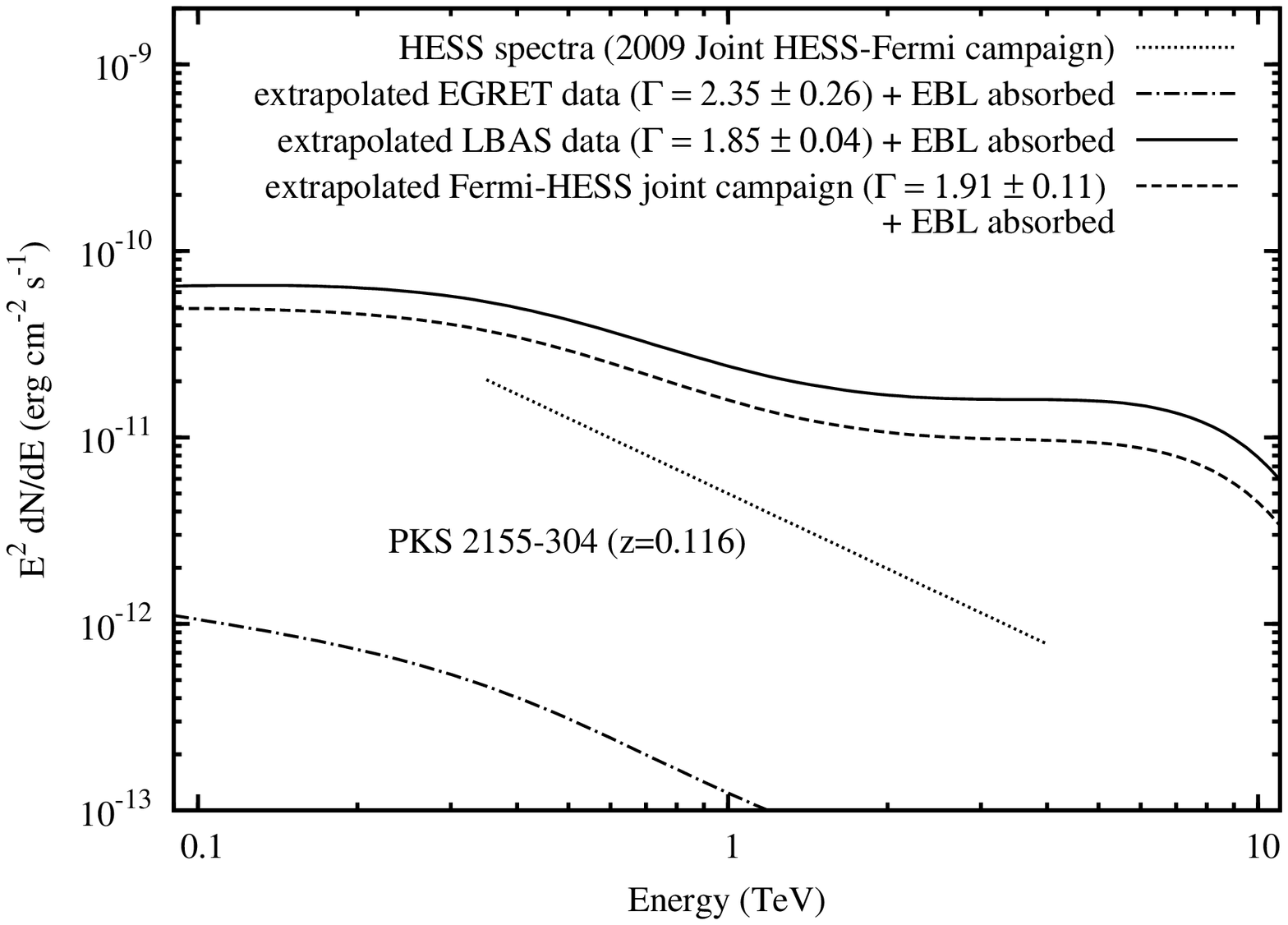}
	  \label{PKS2155-304}}
             }
   \caption{\textbf{\emph{left:}} Comparison of the extrapolated and EBL-absorbed spectra calculated from the FGST spectra as given in the LBAS, is compared to the VHE spectral fit, for the low frequency-peaked blazar, BL Lacertae. The data from LBAS and MAGIC were not simultaneous. \textbf{\emph{right:}} The same for PKS\,2155-304 a high frequency BL Lac object, except that VHE data was taken in an overlapping period in which the LBAS spectra is calculated. The extrapolation from the Fermi data restricted to the strictly simultaneous period is shown with the dashed line.}
   \label{compareToVHE}
\end{figure*}

The Integral flux under the extrapolated spectra, beyond $200$\,GeV, I($>200$\,GeV), gives an indication of the flux that would be seen using the current generation of IACTs such as VERITAS, H.E.S.S., and MAGIC. Assuming that an I($>200\,$GeV) \,$>\approx$\,$10^{-12}$ is required for a $5$\,$\sigma$ detection in $50$\,hours, a list of $20$ candidates satisfying this condition, ranked by their I($>200$\,GeV) value is presented in table \ref{RankedList}. In the table, F$_{100}$ is the LBAS flux in the $0.1$\,GeV to $100$\,GeV band. $\Gamma_{Fermi}$ (also from LBAS) is the spectral index derived from a simple power law fit between $0.2$\,GeV and $100$\,GeV. Established VHE emitting blazars are denoted with an asterisk after their common names. The fact that there are eight VHE emitters in this list of twenty predicted VHE emitters, gives credibility to this selection method.\\

\section{Comparing to actual VHE measurements}
To compare these predictions with actual VHE spectra of the sources, simultaneous observation using IACTs need to be compared with the LBAS extrapolations that we give here. The published example of such a simultaneous observation is the joint H.E.S.S.-Fermi campaign on PKS 2155-304 between $25^{th}$ August $2008$ and $6^{th}$ September $2008$, \cite{PKS2155-JointCampaign}. The FGST spectra measured during the campaign is consistent with the values given in the LBAS. However, the spectra in \cite{PKS2155-JointCampaign} shows a break in the Fermi spectra at $1$\,GeV. Since it is not known to us if this break is a stable feature in the GeV spectra throughout the three month period used to get the LBAS values, we simply take the single power-law fits from LBAS. Given this we expect the extrapolations to VHE regime to lie slightly above the true VHE measurements, provided the correction due to the EBL absorption was correct (assuming that the EBL used here truly represents an upper limit). As we show in figure \ref{PKS2155-304} this is indeed the case, for both the simple power law fit from the LBAS, and the simple power-law fit in \cite{PKS2155-JointCampaign}.\\

Comparing to non-simultaneous VHE datasets though not as constraining to check our estimates, still gives an indication of their compatibility. Two more examples are given in figures \ref{Bl-Lac} and \ref{PKS2155-304}. Such a comparison was done for 10 sources, and an overall good agreement was found between our extrapolations and the true-VHE measurements. It is worthwhile to note that, in \cite{BeheraWagner2008}, extrapolations from the EGRET values consistently under predicted the VHE spectra. This could be due to two possible reasons. Firstly most blazars are likely to show long term variability; and secondly the LAT instrument on Fermi has a higher high-energy threshold, and thus is much nearer to the VHE regime than EGRET, thus yielding extrapolations that are much closer to the true intrinsic VHE spectra.\\

\section{Conclusion}
With the data available from the FGST, it is now possible for the first time to make simultaneous measurements of the high energy component of VHE blazars, from 100 MeV to tens of TeV. It also opens up a window\newpage \noindent of opportunity to discover new VHE blazars based on extrapolations from FGST measurements. A list of such candidates is provided here using the spectral measurements from the LAT bright AGN sample. The extrapolated spectra of these candidates are above the detection threshold of the current generation IACTs, and are recommended for observations. Eight of the twenty listed blazars are established VHE emitters and hence validate this strategy for selecting candidates for VHE observations.

\end{document}